\documentclass[floatfix,twocolumn]{revtex4}
\usepackage{amsmath}
\usepackage{color}
\usepackage{amssymb}
\usepackage{graphicx}
\usepackage[utf8]{inputenc}
\usepackage[english, russian]{babel}
\usepackage{textcomp}
\usepackage{color}
\usepackage{esint}

\begin{document}

\title{ 
Altermagnetic and Noncentrosymmetric Metals}

\author{V.P.Mineev}
\affiliation{Landau Institute for Theoretical Physics, 142432 Chernogolovka, Russia}

\begin{abstract}

 A theory of the normal and superconducting states of piezomagnetic metals, which are altermagnetic materials, has been developed. This has been done in comparison with the corresponding theoretical description for metals that do not have spatial inversion symmetry. Particular attention has been paid to the problem of the anomalous Hall effect, and its absence has been demonstrated in both altermagnetic and non-centrosymmetric metals. It has been shown that the superconducting state in altermagnets always arises by pairing electrons from zones split by spin-orbit interaction, which makes the existence of such superconductors unlikely.

\end{abstract}
\date{\today}
\maketitle

\section{Introduction}
The possibility of the existence of substances with a magnetic crystal structure, the symmetry group of which \\
(i) does not contain the time reversal operation R or contains it only in combination with rotations or reflections, \\
(ii) does not contain the operation R and spatial inversion I separately, but contains their product IR, was noted in the fundamental work of B.A. Tavger and V.M. Zaitsev \cite{Tavger1956}.
 In the first case, this leads to the effect of piezomagnetism, that is, the emergence of magnetization when mechanical stress is applied to the crystal, as well as to the appearance of deformation linear with respect to the magnetic field applied to the crystal \cite{LL}. In the second case, we are dealing with the phenomenon of magnetoelectricity, that is, the appearance of a magnetic moment when an electric field is applied to the crystal and, conversely, to the appearance of electric polarization when a magnetic field is applied \cite{LL}. The first examples of substances possessing piezomagnetic or magnetoelectric properties were indicated by I.E. Dzyaloshinsky \cite{Dzyal1957L, Dzyal1959}.
Soon after, piezomagnetism was discovered in antiferromagnetic fluorides of cobalt CoF$_2$ and manganese MnF$_2$
\cite{Borovik1960}, and the magnetoelectric effect was discovered \cite {Astrov1960} in the antiferromagnet Cr$_2$O$_3$.

The group of symmetry of CoF$_2$ and MnF$_2$ is
\begin{eqnarray}
{\bf D}_{4h}({\bf D}_{2h})=( E, C_{2z}, \sigma_h, 2\sigma_v, 2U_2, I, \nonumber\\
2C_{4z}R, 2U_2^\prime R, 2\sigma_v^\prime R, 2C_{4z}\sigma_h R).
\label{1}
\end{eqnarray}
The piezomagnetic thermodynamic potential invariant in respect of all these operations is
\begin{equation}
\Phi_{pm}=-\lambda_1(\sigma_{xz}H_y+\sigma_{yz}H_x)-\lambda_2\sigma_{xy}H_z
\end{equation}
and corresponding additional magnetisation arising under application of shear stress  $\sigma_{xz}$  is
\begin{equation}
M_y=-\frac{\partial \Phi_{pm}}{\partial H_y}=\lambda_1\sigma_{xz}.
\end{equation}
Both CoF$_2$ and MnF$_2$ are dielectric antiferromagnets. Their magnetic structure is shown on Fig1.

\begin{figure}
\includegraphics
[height=.2\textheight]
{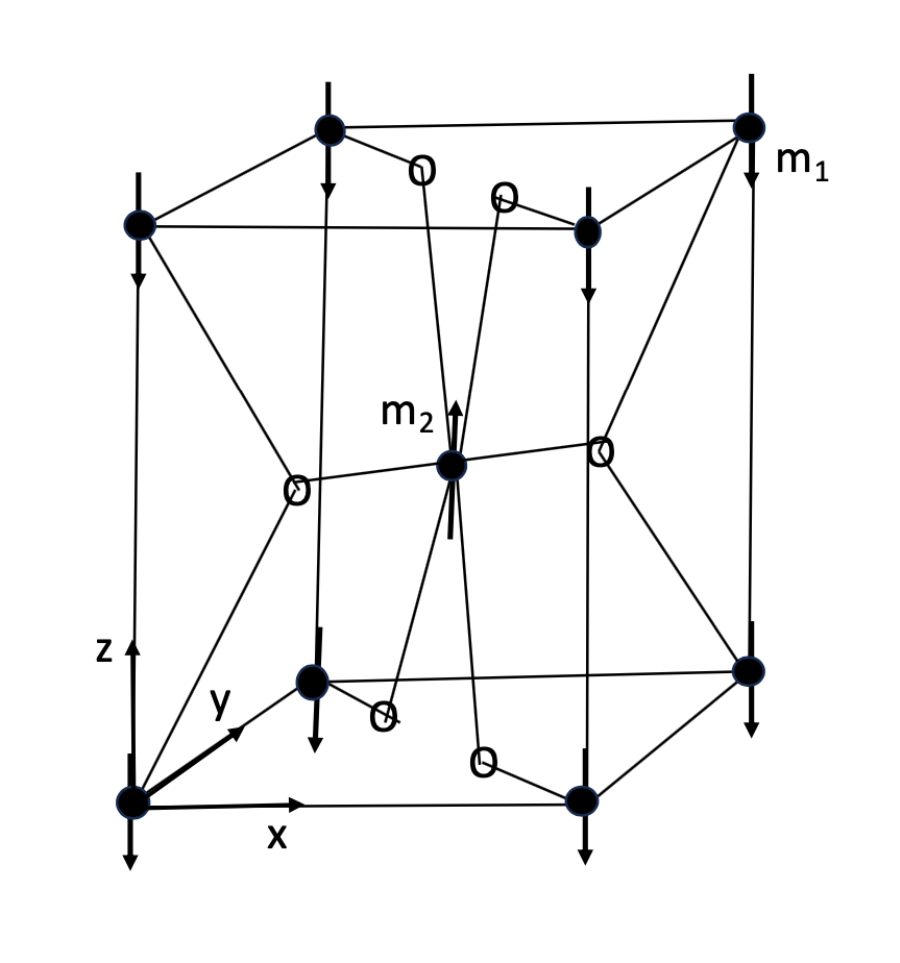}
 \caption{
Magnetic structure of dielectric MnF$_2$ showing the order and orientation of the Mn ions magnetic moments. The small circles correspond to fluorine sites. }
\end{figure}

 Let us now imagine that some metallic material ( RuO$_2~$- ??) possesses by the same symmetry. 
Near the $\Gamma$-point  in reciprocal space the electron spectrum of this metal  invariant in respect of all operation ${\bf D}_{4h}({\bf D}_{2h})$ is
\begin{equation}
\hat\varepsilon({\bf k})=\varepsilon_{\bf k}\sigma_0+\gamma_1(k_xk_z\sigma_y+k_yk_z\sigma_x)+\gamma_2k_xk_y\sigma_z,
\label{alt1}
\end{equation}
where 
$\varepsilon_{\bf k}$ — spin independent part of spectrum, $\sigma_0$ is unit  2 × 2 matrix in the spin space,
and
$\mbox{\boldmath$\sigma$}=(\sigma_x,\sigma_y,\sigma_z)$ are the Pauli matrices. 
In general, the electron spectrum of a metal such that its group of symmetry $G$ ( magnetic class) contains the operation of time reversal only in combination with rotations or reflections has the form 
\begin{equation}
\varepsilon_{\alpha\beta}({\bf k})=\varepsilon_{\bf k}\delta_{\alpha\beta}+\mbox{\boldmath$\gamma$}_{\bf k}\mbox{\boldmath$\sigma$}_{\alpha\beta},
\label{alt}
\end{equation}
invariant in respect of all operations of the group  $G$. There is subclass of  these type metals such that the angular average 
\begin{equation}
\int\frac{d\Omega_{\bf k}}{4\pi}\mbox{\boldmath$\gamma$}_{\bf k}=0.
\label{am1}
\end{equation}
These type of metals looking like antiferromagnets in reciprocal space are called {\bf altermagnets}. 
The electron spin ordering in altermagnets determined by spin-orbit coupling is  in general  non-collinear.

We will consider the properties of metals belonging to the above-defined class. At the same time there is a vast literature devoted to description
of materials called altermagnets with  magnetic order  staggered both in the coordinate space and in the momentum space
with collinear and non-collinear spin-order  performed in terms of particular microscopic models.
See for example \cite{Smejkal2022}.

\bigskip

The magnetoelectric material Cr$_2$O$_3$ point symmetry group
\begin{equation}
{\bf D}_{3d}({\bf D}_3)=(E, C_3,C_3^2,3u_2,3\sigma_dR,2S_6R,IR)
\end{equation} 
 contains the product of time and space inversion, but does not include these operations separately. The corresponding thermodynamic potential invariant in respect to these operations is
 \begin{equation}
 \Phi_{em}=-\alpha_\perp(E_xH_x+E_yH_y)-\alpha_\parallel E_zH_z.
 \end{equation}
So, this material in external electric field acquires magnetisation
\begin{equation}
M_y=-\frac{\partial \Phi_{em}}{\partial H_y}=\alpha_\perp E_y.
\end{equation}
Cr$_2$O$_3$ is antiferromagnetic dialectric. A metal with the same symmetry as Cr$_2$O$_3$ also possesses magnetoelecric properties.  Besides this its electron spectrum invariant in respect of all operations of the group ${\bf D}_{3d}({\bf D}_3)$
\begin{eqnarray}
\varepsilon_{\bf k}=\varepsilon^e_{\bf k}+\varepsilon^o_{\bf k},~~~~~~\varepsilon^e_{\bf k}=f(k_x^2+k_y^2,k_z^2),\\\varepsilon^o_{\bf k}=\gamma
(3k_y^2k_x-k_x^3)~~~~~~~
\label{epsilon}
\end{eqnarray}
consists from two parts even and odd in respect to  its argument ${\bf k}$.  
This is a general property of a metal with a symmetry that does not include the operations of time inversion $R$ and space inversion $I$ separately, but is invariant with respect to its product $IR$. Such metals are called {\bf metals with toroidal order} or simply {\bf toroids}. Normal properties and superconducting states in toroids were discussed in the article \cite{Mineev2024}.

This article is devoted to the description of the properties of altermagnets in the normal and superconducting states.
This is conveniently done by comparing altermagnets and non-centrosymmetric metals. We begin with a description of the electron states in altermagnets. Then, using the kinetic equation, we discuss the problem of anomalous Hall effect in altermagnets in the absence of external magnetic field. Finally, a formal description of superconductivity is given.

 \section{Electronic states and magnetic moment}
 
 The electron spectrum in altermagnets (\ref{alt}) has the same form as in noncentrosymmetric metals 
\begin{equation}
\hat\varepsilon({\bf k})=\varepsilon_{\bf k}\sigma_0+\mbox{\boldmath$\gamma$}_{\bf k}\mbox{\boldmath$\sigma$}.
\label{11}
\end{equation} 
See for example \cite{Samokhin2008} and references therein.
Thus all the calculations for these different types of metals look identical, but one must remember that in altermagnets vector 
$
\mbox{\boldmath$\gamma$}_{-{\bf k}}=\mbox{\boldmath$\gamma$}_{\bf k}
$
is even function of ${\bf k}$, whereas in noncentrosymmetric metals it is odd one $
\mbox{\boldmath$\gamma$}_{-{\bf k}}=-\mbox{\boldmath$\gamma$}_{\bf k}
$. Scalar part of spectrum $\varepsilon_{\bf k}=\varepsilon_{-{\bf k}}$ is even in both cases.

The eigenvalues of the matrix (\ref{11}) are
\begin{equation}
    \varepsilon_{+}({\bf k})=\varepsilon+\gamma,~~~~~~~~ \varepsilon_{-}({\bf k})=\varepsilon-\gamma,
\label{e3}
\end{equation}
where $\gamma=|\mbox{\boldmath$\gamma$}_{\bf k}|$.
The corresponding eigenfunctions are given by
\begin{eqnarray}
\Psi^+_\alpha({\bf k})=\frac{1}{\sqrt{2\gamma(\gamma+\gamma_z)}}\left (\begin{array} {c}
\gamma+\gamma_z\\
\gamma_+
\end{array}\right),\nonumber\\
~~~~~~~~~~~~\Psi^-_\alpha({\bf k})=\frac{t_+^\star}{\sqrt{2\gamma(\gamma+\gamma_z)}}
\left(\begin{array} {c}
-\gamma_-\\
\gamma+\gamma_z
\end{array}\right),
\label{ps}
\end{eqnarray}
where $\gamma_\pm=\gamma_x\pm i\gamma_y$ and $t_+^\star=-\frac{\gamma_+}{\sqrt{\gamma_+\gamma_-}}$.
The eigenfunctions obey the orthogonality conditions
\begin{equation}
\Psi^{\lambda_1\star}_\alpha({\bf k})\Psi^{\lambda_2}_\alpha({\bf k})=\delta_{\lambda_1\lambda_2},~~~~~~~
\Psi^\lambda_{\alpha}({\bf k})\Psi^{\lambda\star}_{\beta}({\bf k})=\delta_{\alpha\beta}.
\label{ort}
\end{equation}
Here, a summation over the repeating  spin $\alpha=\uparrow,\downarrow$
or band $\lambda=+,-$ indices is implied. 

In altermagnets as in noncentrosymmetric metals the eigen functions are related to each other by operation of time inversion $-i(\sigma_y)K_0$, where $K_0$ is the operation of complex conjugation,
$$-i(\sigma_y)_{\alpha\beta}K_0\Psi_{\beta}^+({\bf k})\propto\Psi_{\alpha}^-({\bf k}).$$  Thus, the Kramers degeneracy is lifted.

There are two Fermi surfaces determined by the equations
\begin{equation}
\label{e4}
    \varepsilon_{+}({\bf k})=\mu,~~~~~~~~~~~~\varepsilon_{-}({\bf k})=\mu
\end{equation}
and the Fermi velocities are given by the derivatives
\begin{equation}
{\bf v}_{+}=\frac{\partial\varepsilon_{+}({\bf k})}{\partial {\bf k}},~~~~~~~~~~~~~~{\bf v}_{-}=\frac{\partial\varepsilon_{-}({\bf k})}{\partial {\bf k}}.~
\end{equation}

The spin quantisation axis is  given by the unit vector $\hat {\mbox{\boldmath$\gamma$}}=\mbox{\boldmath$\gamma$}/|\mbox{\boldmath$\gamma$}|$.
The  projections of the electron spins in two bands on the $\hat {\mbox{\boldmath$\gamma$}}$ direction have opposite orientations 
\begin{equation}
(\hat {\mbox{\boldmath$\gamma$}}_{\bf k}\mbox{\boldmath$\sigma$}_{\alpha\beta})\Psi^{\pm}_{\beta}({\bf k})=\pm\Psi^{\pm}_{\alpha}({\bf k}).
\end{equation}

In an external magnetic field the matrix of the electron  energy is
\begin{equation}
\hat\varepsilon({\bf k})=\varepsilon_{\bf k}\sigma_0+\mbox{\boldmath$\gamma$}_{\bf k}\mbox{\boldmath$\sigma$}-{\bf h}\mbox{\boldmath$\sigma$}.
\label{hmatrix}
\end{equation}
The field is here written as ${\bf h}=\mu_B{\bf H}$.
The band energies are now given by
\begin{equation}
 \varepsilon_{\lambda,{\bf h}}({\bf k})=\varepsilon_{\bf k}+\lambda
|\mbox{\boldmath$\gamma$}_{\bf k}-{\bf h}|, ~~~\lambda=\pm.
\end{equation}
Along with the changes of the band energies, the spin quantisation axis also deviates from its zero field direction
\begin{equation}
\hat {\mbox{\boldmath$\gamma$}}_{\bf k}~~~\to~~~\hat {\mbox{\boldmath$\gamma$}}_{\bf h}({\bf k})=
\frac{\mbox{\boldmath$\gamma$}_{\bf k}-{\bf h}}{|\mbox{\boldmath$\gamma$}_{\bf k}-{\bf h}|}.
\end{equation}
The magnetic moment is written as
\begin{equation}
{\bf M}=\mu_B\int \frac{d^3k}{(2\pi)^3}\hat {\mbox{\boldmath$\gamma$}}_{\bf h}({\bf k})\left[n( \varepsilon_{+,{\bf h}}({\bf k}))
-n( \varepsilon_{-,{\bf h}}({\bf k}))\right],
\end{equation}
where $n(\varepsilon_\lambda)=\left(e^{\frac{\varepsilon_\lambda-\mu}{T}}+1\right)^{-1}$ is the Fermi distribution function.
Taking the term of the first order in magnetic field
we obtain for the magnetic susceptibility
\begin{eqnarray}
\chi_{ij}=-\mu_B^2\int \frac{d^3k}{(2\pi)^3}\left\{\hat {\mbox{\boldmath$\gamma$}}_i\hat {\mbox{\boldmath$\gamma$}}_j
\left[\frac{\partial n( \varepsilon_{+})}{\partial\varepsilon_+}
+\frac{\partial n( \varepsilon_{-})}{\partial\varepsilon_-}\right]+\right.\nonumber\\
\left.+(\delta_{ij}-\hat {\mbox{\boldmath$\gamma$}}_i\hat {\mbox{\boldmath$\gamma$}}_j )\frac{n( \varepsilon_{+})
-n( \varepsilon_{-})}{|\mbox{\boldmath$\gamma$}|}\right\}.~~~~~~~~
\label{chi}
\end{eqnarray}
The first term under the sign of integration contains the derivatives of the jumps in the Fermi distributions ${\partial n( \varepsilon_{\pm})}/{\partial\varepsilon_{\pm}}=-\delta(\varepsilon_\pm-\mu)$. The second one originates from the deviation in the spin quantisation direction for the quasiparticles filling the states between the Fermi surfaces of two bands.
Thus, magnetic moment arising  in altermagnets in external magnetic field is determined by the same formula as in noncentrosymmetric metals  \cite{Mineev2010}. 

Altermagnets are invariant in respect of space inversion, hence the external electric field does not cause magnetisation to appear in them. On the contrary noncentrosymmetric metals placed in an electric field possess  magnetolectric effect $M_i=A_{ij}E_j=A_{ij}\rho_{jk}j_k$, where ${\bf j}$ is density of dissipative current \cite{MineevLT2024}.

\section{Kinetic equation and Hall effect}

The matrix of the equilibrium electron distribution function is
\begin{equation}
\hat n
=\frac{n({\varepsilon}_+)+n({\varepsilon}_-)}{2}\sigma_0+\frac{n({\varepsilon}_+)-n({\varepsilon}_-)}{2\gamma} 
\mbox{\boldmath$\gamma$}\cdot \mbox{\boldmath$\sigma$}.
\label{eqv}
\end{equation}
In the band representation the equilibrium distribution function  is given by the diagonal matrix
\begin{equation}
n_{\lambda_1\lambda_2}=\Psi^{\lambda_1\star}_{\alpha}({\bf k})n_{\alpha\beta}\Psi^{\lambda_2}_{\beta}({\bf k})=\left (\begin{array} {cc} n({\varepsilon}_+)&0\\0&n({\varepsilon}_-)  \end{array}\right)_{\lambda_1\lambda_2}.
\end{equation}

The Hermitian matrices of the non-equilibrium distribution functions in the band and spin representations  related by
\begin{equation}
f_{\lambda_1\lambda_2}({\bf k})=\Psi^{\lambda_1\star}_{\alpha}({\bf k})f_{\alpha\beta}\Psi^{\lambda_2}_{\beta}({\bf k}).
\label{f}
\end{equation}

The kinetic equation for the electron distribution function in non-centrosymmetric metals has been obtained in \cite{Mineev2019}  from  the general matrix quasi-classic kinetic equation derived by V.P.Silin \cite{Silin1957}. 
In presence of time dependent  electric field ${\bf E}(t)={\bf E}_{\omega}e^{-i\omega t}$ the 
the linearised matrix  kinetic equation  for the  frequency dependent  Fourier amplitudes of deviation of distribution function from equilibrium  $g_{\lambda_1\lambda_2}({\bf k},\omega)$ 
is
\begin{eqnarray}
-i\omega
 \left (\begin{array} {cc}g_+&g_{\pm}\\g_{\mp}&g_-
 \end{array}\right)~~~~~~~~~~~~~~~~~~~\nonumber\\
 +e\left (\begin{array} {cc}({\bf v}_{+}{\bf E}) \frac{\partial n_+}{\partial \varepsilon_+}&({\bf w}_{\pm}{\bf E})(n_--n_+)
\\
({\bf w}_{\mp}{\bf E})(n_+-n_-))&({\bf v}_{-}{\bf E})\frac{\partial n_-}{\partial \varepsilon_-}
 \end{array}\right)~~~~~~~\nonumber\\+
 \left(
\begin{array} {cc}0&i(\varepsilon_--\varepsilon_+)g_{\pm}({\bf k})\\
i(\varepsilon_+-\varepsilon_-)g_{\mp}({\bf k})&0
 \end{array}\right)=\hat I.~~
 \label{eqv1}
\end{eqnarray}
Here, we put for brevity $n(\varepsilon_+)=n_+, ~n(\varepsilon_-)=n_-$. The quantities
\begin{eqnarray}
{\bf w}_{\pm}({\bf k})=
\Psi^{+\star}_{\alpha}({\bf k})\frac{\partial \Psi^{-}_{\alpha}({\bf k})}{\partial{\bf k}}~~~~~~~~\nonumber\\=\frac{t_+^\star}{2\gamma}\left (-\frac{\partial \gamma_-}{\partial{\bf k}}+\frac{\gamma_-}{\gamma+\gamma_z}\frac{\partial(\gamma+\gamma_z)}{\partial{\bf k}}\right),~~~~~
\label{vel}
\end{eqnarray}
are {\bf the interband} Berry connections,
$$
{\bf w}_{\mp}=-{\bf w}_{\pm}^\star.
$$
 Unlike to the group velocities ${\bf v}_+$,  ${\bf v}_-$, the dimensionality of the Berry connections ${\bf w}_\pm$
and ${\bf w}_\mp$ is $1/k$. 
$\hat I$ is the matrix integral of scattering. It has quite cumbersome form even in the case of scattering on impurities in the Born approximation \cite{Mineev2019}.

We will consider collisionless limit and put $\hat I=0$. Then 
the solutions of Eq.(\ref{eqv1}) are:
\begin{eqnarray}
g_+=\frac{e}{i\omega}({\bf v}_{+}{\bf E}) \frac{\partial n_+}{\partial \varepsilon_+},
\label{A}\\
g_-=\frac{e}{i\omega}({\bf v}_{-}{\bf E}) \frac{\partial n_-}{\partial \varepsilon_-},
\label{B}\\
g_{\pm}=\frac{e({\bf w}_{\pm}{\bf E})(n_--n_+)}{i(\omega-\varepsilon_-+\varepsilon_+)},
\label{C}\\
g_{\mp}=\frac{e({\bf w}_{\mp}{\bf E})(n_+-n_-)}{i(\omega-\varepsilon_++\varepsilon_-)}.
\label{D}
\end{eqnarray}

  The electric  current density is
 \begin{equation}
 {\bf j}=e\int \frac{d^3k}{(2\pi)^3}\frac{\partial \varepsilon_{\alpha\beta}({\bf k})}{\partial {\bf k}}g_{\beta\alpha}({\bf k},\omega).
 \end{equation}
Transforming it to the band representation we obtain
\begin{eqnarray}
 {\bf j}=e\int \frac{d^3k}{(2\pi)^3} \Psi_\alpha^{\lambda_1\star}({\bf k})\frac{\partial \varepsilon_{\alpha\beta}({\bf k})}{\partial {\bf k}}
 \Psi_{\beta}^{\lambda_2}({\bf k})\Psi_{\gamma}^{\lambda_2\star}({\bf k})
 g_{\gamma\delta}({\bf k},\omega)\Psi_{\delta}^{\lambda_1}({\bf k})\nonumber\\
   =e\int \frac{d^3k}{(2\pi)^3} \left \{
 \frac{\partial \varepsilon_{\lambda_1\lambda_2}({\bf k})}{\partial {\bf k}}+\left [ {\bf w}_{\lambda_1\lambda_3},  \varepsilon_{\lambda_3\lambda_2} \right]\right\}
 g_{\lambda_2\lambda_1}({\bf k})
,~~
\label{current}
\end{eqnarray}
where $\left [\dots,\dots\right ]$ is the commutator.
 Performing matrix multiplication we obtain
\begin{equation}
{\bf j}=e\int \frac{d^3k}{(2\pi)^3} \left [ {\bf v}_+g_++{\bf v}_-g_-+({\bf w}_{\pm}g_{\mp}- {\bf w}_{\mp}g_{\pm})(\varepsilon_--\varepsilon_+)  \right ].
\label{cur}
\end{equation}
Substitution Eqs.(\ref{A})-(\ref{D}) to the Eq.(\ref{cur}) gives
\begin{eqnarray}
 {\bf j}=e^2\int \frac{d^3k}{(2\pi)^3} \left\{
\frac{{\bf v}_+({\bf v}_{+}{\bf E})}{i\omega} \frac{\partial n_+}{\partial \varepsilon_+}
+
\frac{{\bf v}_-({\bf v}_{-}{\bf E})}{i\omega} \frac{\partial n_-}{\partial \varepsilon_-}
\right.\nonumber\\
\left.+\frac{(n_+-n_-)(\varepsilon_--\varepsilon_+)}{\omega^2-(\varepsilon_+-\varepsilon_-)^2}
\left [2i\omega  Re({\bf w}_\pm({\bf w}_\pm^\star{\bf E}))\right.\right.\nonumber\\
+\left.\left.(\varepsilon_--\varepsilon_+)]2 Im({\bf w}_\pm({\bf w}_\pm^\star{\bf E}))\right ]
 \right \}.
\label{cur1}
\end{eqnarray}
The last term in this formula contains combination
\begin{equation}
2 Im\left (w_\pm)_i(w_\pm^\star)_jE_j\right)=
\frac{\mbox{\boldmath$\gamma$}}{\gamma^3}
\left(\frac{\partial \mbox{\boldmath$\gamma$}}{\partial k_i}\times
  \frac{\partial \mbox{\boldmath$\gamma$}}{\partial k_j}   \right  )E_j=\Omega_{ij}({\bf k})E_j
\end{equation}
including the antisymmetric tensor 
 \begin{equation}
 \Omega_{ij}({\bf k})=-\Omega_{ji}({\bf k}).
 \end{equation}
 
Thus, one can expect the occurrence an electric potential in the direction perpendicular to the current, i.e.
the Hall effect in the absence of an external magnetic field. 
 However,
in noncentrosymmetric metals $\mbox{\boldmath$\gamma$}_{-{\bf k}}=-\mbox{\boldmath$\gamma$}_{{\bf k}}$ and $\Omega_{ij}(-{\bf k})=-\Omega_{ij}({\bf k})$.
Hence, the last term in Eq.(\ref{cur1}) is vanishing. In altermagnets 
 $\mbox{\boldmath$\gamma$}_{-{\bf k}}=\mbox{\boldmath$\gamma$}_{{\bf k}}$ 
 and $\Omega_{ij}({\bf k})$ is even but sign alternating function such that
$\int\frac{d\Omega_{\bf k}}{4\pi}\Omega_{ij}({\bf k})=0$ and the last term in Eq.(\ref{cur1}) is vanishing as well. This is easy to check
 for instance in particular case of altermagnet with symmetry ${\bf D}_{4h}({\bf D}_{2h}) $ with $\mbox{\boldmath$\gamma$}_{{\bf k}}$  given by Eq.(\ref{alt1}).
Thus, in  altermagnets the anomalous Hall effect in absence of an external field is absent. 
This conclusion is truth for substances not possessing  spontaneous magnetization that is under fulfilment condition (\ref{am1}). By much more cumbersome calculations one can 
 verify the validity of this statement also  taking into account scattering processes when $\hat I\ne 0$.

In the absence of  limitation (\ref{am1})
as it is for instance  in metal  with symmetry group ${\bf D}_{2h}({\bf C}_{2})=(E, C_{2z}, RC_{2x},RC_{2y}) $ 
\begin{equation}
\mbox{\boldmath$\gamma$}_{{\bf k}}=\gamma_xk_xk_z\hat x+\gamma_yk_yk_z\hat y+\gamma_zk_z^2\hat z
\label{g}
\end{equation}
and 
\begin{equation}
\Omega_{xy}({\bf k})=\frac {\gamma_x\gamma_y\gamma_zk_z^4}{[ (\gamma_xk_xk_z)^2+ (\gamma_yk_yk_z)^2+(\gamma_zk_z^2)^2  ) ^{3/2}  }.
\end{equation}
Thus,  the last term in Eq.(\ref{cur1}) has the finite value. Hence,  the anomalous Hall effect in absence of an external field takes place. This is not astonishing, however, because at $\int\frac{d\Omega_{\bf k}}{4\pi}\mbox{\boldmath$\gamma$}_{{\bf k}}\ne 0$ the substance possesses spontaneous magnetisation.

\section{Superconducting states }

The  BCS  Hamiltonian for singlet pairing in the  spinor basis has the following form 
\begin{eqnarray}
    \hat H=\sum_{{\bf k}\alpha\beta}(\varepsilon_{\bf k}\delta_{\alpha\beta}+\mbox{\boldmath$\gamma$}_{\bf k}\mbox{\boldmath$\sigma$}_{\alpha\beta}-\mu\delta_{\alpha\beta})
   a^\dagger_{{\bf k}\alpha}a_{{\bf k}\beta} ~~~~~\nonumber\\+  
    \frac{1}{2}\sum\limits_{{\bf k}{\bf k}'}\sum_{\alpha\beta\gamma\delta}
  V({\bf k},{\bf k}')(i\sigma_y)_{\alpha\beta}(\sigma_y)^\dagger_{\gamma\delta} 
      a^\dagger_{-{\bf k}\alpha}a^\dagger_{{\bf k}\beta}a_{{\bf k}'\gamma}a_{-{\bf k}'\delta}.~~
\end{eqnarray}
Here
\begin{equation}
V({\bf k},{\bf k}')=-V_0\varphi^\Gamma_i({\bf k})\varphi^{\Gamma\star}_i({\bf k}'),
\end{equation}
is the paring potential decomposed over basis of even $\varphi^\Gamma_i({\bf k})=\varphi^\Gamma_i(-{\bf k})$  functions of given irreducible representation $\Gamma$ of the crystal symmetry group. For example, for altermagnet with symmetry group ${\bf D}_{4h}({\bf D}_{2h})$ consisting of operations enumerated in Eq.(\ref{1}) the
function transforming according one-dimensional unit representation is 
\begin{equation}
\varphi({\bf k})\propto~i(\hat k_x^2-\hat k_y^2).
\end{equation}
Here $\hat k_x,\hat k_y$ are the components of unit vector ${\bf k}/k_F$.
Transforming  to the band representation
\begin{equation}
a_{{\bf k}\alpha}=\Psi^\lambda_\alpha({\bf k})c_{{\bf k}\lambda}
\end{equation}
we obtain
\begin{eqnarray}
\hat H=\sum_{{\bf k}\lambda}(\varepsilon_{\lambda}({\bf k})-\mu)
   c^\dagger_{{\bf k}\lambda}c_{{\bf k}\lambda} ~~~~~~~~~~~\nonumber\\+
   \frac{1}{2}\sum\limits_{{\bf k}{\bf k}'}\sum_
   {\lambda_1\lambda_2\lambda_3\lambda_4}
V_{\lambda_1\lambda_2\lambda_3\lambda_4}({\bf k},{\bf k}') 
      c^\dagger_{-{\bf k}\lambda_1}c^\dagger_{{\bf k}\lambda_2}c_{{\bf k}'\lambda_3}c_{-{\bf k}'\lambda_4},~~
\end{eqnarray}
\begin{equation}
V_{\lambda_1\lambda_2\lambda_3\lambda_4}({\bf k},{\bf k}')=V({\bf k},{\bf k}')t_{\lambda_2}({\bf k})t_{\lambda_4}^\star({\bf k}^\prime)\sigma^x_{\lambda_1\lambda_2}\sigma^x_{\lambda_3\lambda_4} ~~~~
\end{equation}
where $t_\lambda({\bf k})=-\lambda\frac{\gamma_-}{\sqrt{\gamma_+\gamma_-}}$ is the phase factor. It is obvious from  this expression that pairing in altermagnets is the pairing of electrons from different bands. This distinguishes them from noncentrosymmetric metals where the pairing  mostly occurs between the electrons from the same band \cite{Samokhin2008}. The situation in altermagnets reminds pairing in conventional superconductors with singlet pairing in magnetic field which splits the Fermi surfaces with opposite spins. That leads to paramagnetic suppression of superconductivity. In altermagnets  the same effect takes place in a field absence that leads to effective reduction of temperature of transition to superconducting state or even to complete suppression of superconductivity. Thus, the possibility of existence of superconducting altermagnets raise doubts. Nevertheless, for completeness we present here the  theoretical description of superconductivity in altermagnets.

The Gor'kov equations are
\begin{eqnarray}
 \begin{pmatrix} i\omega\delta_{\lambda_1\lambda_2}-H_{\lambda_1\lambda_2}&-\tilde\Delta_{\lambda_1\lambda_2} \\
 -\tilde\Delta^\dagger_{\lambda_1\lambda_2}& i\omega\delta_{\lambda_1\lambda_2}+H_{\lambda_1\lambda_2}\end{pmatrix}
 \begin{pmatrix}G_{\lambda_2\lambda_3}& -\tilde F_{\lambda_2\lambda_3}\\
-\tilde F^\dagger_{\lambda_2\lambda_3}&-G_{\lambda_2\Lambda_3 }\end{pmatrix}\nonumber\\
=\delta_{\lambda_1\lambda_3} \begin{pmatrix}1&0\\0&1\end{pmatrix},~~~~~
\end{eqnarray} 
where 
\begin{equation}
 i\omega\delta_{\lambda_1\lambda_2}-H_{\lambda_1\lambda_2}=\begin{pmatrix}i\omega-\varepsilon_++\mu&0\\0&i\omega+\varepsilon_--\mu\end{pmatrix}
\end{equation}
and the phase factor is absorbed in the expressions for the order parameter and the Gor'kov function:
\begin{equation}
\Delta_{\lambda_1\lambda_2}({\bf k})=t_{\lambda_2}({\bf k})\tilde\Delta_{\lambda_1\lambda_2}({\bf k}),
\end{equation}
\begin{equation}
\tilde\Delta_{\lambda_1\lambda_2}({\bf k})=(\sigma_x)_{\lambda_1\lambda_2}\Delta({\bf k}),
\end{equation}
\begin{equation}
F_{\lambda_1\lambda_2}({\bf k},\omega_n)=
t_{\lambda_2}({\bf k})\tilde F_{\lambda_1\lambda_2}({\bf k},\omega_n),
\end{equation}
where $\omega_n=\pi T(2n+1)$ is the Matsubara frequency.
The self-consistency equation is
\begin{eqnarray}
\tilde\Delta_{\lambda_1\lambda_2}({\bf k})~~~~~~~~~~~~~~~~~~~~~~~\nonumber\\=-\frac{T}{2}\sum_n\sum_{{\bf k}^\prime}V({\bf k},{\bf k}')(\sigma_x)_{\lambda_2\lambda_1}(\sigma_x)_{\lambda_3\lambda_4}\tilde F_{\lambda_3\lambda_4}({\bf k}^\prime,\omega_n),~~
\end{eqnarray}
where 
\begin{eqnarray}
\tilde F_{\lambda_1\lambda_2}({\bf k},\omega_n)~~~~~~~~~~~~~~~~~~~~~~~~~~~~~~\nonumber\\= \Delta\begin{pmatrix}0&G_+^n({\bf k},\omega_n)G_-({\bf k},-\omega_n)\\
 G^n_-({\bf k},\omega_n)G_+({\bf k},-\omega_n)&0
 \end{pmatrix}~~~
\end{eqnarray}
is the matrix Gor'kov function and
\begin{equation}
G_\pm({\bf k},\omega_n)=-\frac{i\omega_n+\varepsilon_\pm-\mu}{\omega_n^2+(\varepsilon_\pm-\mu)^2+\Delta^2},
\end{equation}
\begin{equation}
G^n_\pm({\bf k},\omega_n)=\frac{1}{i\omega_n-\varepsilon_\pm+\mu}
\end{equation}
are the band Green functions  in superconducting and normal state correspondingly.
The order parameter in the spin and band representations are related to each other as
\begin{equation}
\Delta_{\alpha\beta}({\bf k})=(i\sigma_y)_{\alpha\beta}\Delta({\bf k}).
\end{equation}

\section{Conclusion}

The article presents a general theoretical approach to the description of the normal and superconducting states of piezomagnetic metallic altermagnets. Being non-invariant under time inversion or possessing time reversal symmetry
only in combination with rotations or reflections, these materials have electron states split into two bands
which makes their mathematical description similar to the corresponding description of metals that do not have spatial inversion symmetry.
The problem of the anomalous Hall effect is considered and it is shown that in the absence of spontaneous magnetization, the effect is absent both in altermagnets and in non-centrosymmetric metals. It is shown that superconducting states in altermagnets are inevitably built by interband pairing of electrons from bands split by spin-orbit coupling.
Therefore, the prospects for the existence of superconducting altermagnets are not optimistic.

 \bigskip

  FUNDING This work was supported by ongoing institutional funding. No additional grants to carry out or direct this particular research were obtained. 
  
\bigskip 
 
 CONFLICT OF INTEREST The author of this work declares that he has no conflicts of interest.

\end{document}